\documentclass[12pt]{article}
\usepackage{epsf}
\def\beq{\begin{equation}}
\def\eeq{\end{equation}}
\def\ap#1#2#3 {Ann. Phys. (NY) {\bf#1} (19#2) #3}
\def\err#1#2#3 {{\it Erratum} {\bf#1} (19#2) #3}
\def\ib#1#2#3 {{\it ibid.} {\bf#1} (19#2) #3}
\def\ijmp#1#2#3 {Int. J. Mod. Phys. {\bf#1} (19#2) #3}
\def\jetp#1#2#3 {JETP Lett. {\bf#1} (19#2) #3}
\def\mpl#1#2#3 {Mod. Phys. Lett. {\bf#1} (19#2) #3}
\def\np#1#2#3 {Nucl. Phys. {\bf#1} (19#2) #3}
\def\pl#1#2#3 {Phys. Lett. {\bf#1} (19#2) #3}
\def\prep#1#2#3 {Phys. Rep. {\bf#1} (19#2) #3}
\def\prev#1#2#3 {Phys. Rev. {\bf#1} (19#2) #3}
\def\prl#1#2#3 {Phys. Rev. Lett. {\bf#1} (19#2) #3}
\def\sjnp#1#2#3 {Sov. J. Nucl. Phys. {\bf#1} (19#2) #3}
\def\spj#1#2#3 {Sov. Phys. JETP {\bf#1} (19#2) #3}
\def\spu#1#2#3 {Sov. Phys. Usp. {\bf#1} (19#2) #3}
\def\zp#1#2#3 {Zeit. Phys. {\bf#1} (19#2) #3}

\begin{document}
\begin{titlepage}
\begin{center}
{\Large \bf Theoretical Physics Institute \\
University of Minnesota \\}  \end{center}
\vspace{0.2in}
\begin{flushright}
TPI-MINN-00/04-T \\
UMN-TH-1837-00 \\
hep-ph/0001057\\
January 2000 \\
\end{flushright}
\vspace{0.3in}
\begin{center}
{\Large \bf  Weak decays $\Xi_Q \to \Lambda_Q \, \pi$
\\}
\vspace{0.2in}
{\bf M.B. Voloshin  \\ }
Theoretical Physics Institute, University of Minnesota, Minneapolis,
MN
55455 \\ and \\
Institute of Theoretical and Experimental Physics, Moscow, 117259
\\[0.2in]
\end{center}

\begin{abstract}

The weak decays $\Xi_b \to \Lambda_b \, \pi$ and $\Xi_c \to \Lambda_c \,
\pi$, in which the heavy quark is not destroyed, are discussed. The
branching fractions for these decays, corresponding to an absolute rate
of order $0.01 \, ps^{-1}$, should be at a one percent level for the $b$
hyperons and at a (few) per mill level for $\Xi_c$, possibly making
feasible their experimental study in future. It is shown, through an
application of the heavy quark limit, the flavor SU(3) symmetry,  and
PCAC, that the $\Delta I = 1/2$ rule should hold very well in these
decays, and also that the $\Xi_b$ decays are purely $S$ wave in the
symmetry limit, while the difference between the $S$ wave amplitudes of
the $\Xi_c$ decays and those for the $\Xi_b$ is related, in terms of the
heavy quark expansion,  to the difference of the total decay rates
within the ($\Xi_c, \, \Lambda_c$) triplet of charmed hyperons. We also
comment on the amplitudes of the semileptonic transitions $\Xi_Q \to
\Lambda_Q \, \ell \, \nu$ and on the weak radiative decays $\Xi_Q \to
\Lambda_Q \, \gamma$.
\end{abstract}

%\noindent
%PACS: 12.39.Hg, 14.20.Lq, 14.20.Mr, 13.30.-a

\end{titlepage}

The absolutely dominant processes in weak decays of the $b$ and $c$
hadrons are, naturally, those associated with the decay of the heavy
quark. It is with these decays where, understandably, lies the main
interest of the current phenomenological studies, with the prospects for
precision determination of the CKM mixing matrix and for uncovering a CP
violation in $B$ mesons. The present paper however deals with a rather
sub-dominant type of decay of strange heavy hyperons into non-strange
ones, a process closely analogous to the decays of ordinary strange
hyperons, and associated with the decay of the strange, rather than
heavy, quark. These decays are interesting for at least two reasons: one
is that the branching fractions of the decays $\Xi_b \to \Lambda_b \,
\pi$ and $\Xi_c \to \Lambda_c \, \pi$ are not hopelessly small and one
may expect that these decays can be studied experimentally, inasmuch as
it will be feasible to study any exclusive nonleptonic decays of the $b$
and $c$ cascade hyperons\footnote{The branching ratio $B(\Xi_b^- \to
\Lambda_b \, \pi^-)$ may well exceed 1\%, thus possibly being the
largest among the branching fractions for individual exclusive
nonleptonic decay channels of the $\Xi_b^-$.}, and the other reason
being that these decays provide a case for a study of the `old' physics
in a new setting, namely a study of the structure of baryons containing
one heavy quark. Thus these decays offer a testing ground for a
combination of the `older' methods, such as the flavor SU(3) symmetry
and the current algebra with the `newer' theoretical ideas related to
the heavy quark limit. Moreover, as will be shown, the difference
between the amplitudes of the decays $\Xi_c \to \Lambda_c \, \pi$  and
$\Xi_b \to \Lambda_b \, \pi$ is related, through PCAC and the SU(3)
symmetry, to the matrix elements of four-quark operators, that govern,
within the heavy quark expansion, the differences of the total inclusive
weak decay rates within the triplets of the heavy baryons: ($\Xi_c, \,
\Lambda_c$) and ($\Xi_b, \, \Lambda_b$). The latter matrix elements are
a crucial ingredient in understanding the pre-asymptotic effects in the
inclusive decay rates of heavy baryons, which are discussed with a
recently renewed interest in relation to the data \cite{pdg} on
$\tau(\Lambda_b)/\tau(B^0)$ (for a most recent mini-review see
Ref.\cite{bigi}, see also the recent papers \cite{gms,mv}).

A very approximate estimate of the absolute rates of the discussed
decays can be done by comparing them to similar strangeness-changing
decays of ordinary strange hyperons, with rates typically of order $0.01
ps^{-1}$. For the charmed hyperons the mass difference between $\Xi_c$
and $\Lambda_c$ is known \cite{pdg} to be about 180 MeV, i.e. quite
close to the mass differences of ordinary hyperons differing by one unit
of strangeness, and the mass difference between $\Xi_b$ and $\Lambda_b$
should be very close to that for the charmed hyperons due to the heavy
quark limit considerations:
\beq
M(\Xi_b)-M(\Lambda_b)=M(\Xi_c)-M(\Lambda_c) + O(m_c^{-2}-m_b^{-2}),
\label{massdiff}
\eeq
since there are no terms of order $1/m_Q$ in this mass splitting. In a
more detailed consideration, to be discussed in this paper, the baryonic
matrix elements, determining the decay amplitudes, are somewhat
different from those for decays of ordinary hyperon decays, thus the
specific absolute rates can differ from the simplistic estimate.
However, using the mentioned relation of the difference of the
amplitudes for $\Xi_c$ and $\Xi_b$ decays to the difference of the
lifetimes of the two $\Xi_c$ hyperons and the $\Lambda_c$, we find that
this difference alone would correspond to the rate of, e.g. the decay
$\Xi_c^0 \to \Lambda_c \, \pi^{-}$, equal to $9 \times 10^{-3} \,
ps^{-1}$, i.e. in the same range as the simplistic estimate. Thus
comparing these estimates for the absolute rates with the known
lifetimes of the $\Xi_c$ and $\Xi_b$ hyperons, one concludes that the
discussed decays should have branching fractions at a per mill level for
the $\Xi_c$ hyperons and at a one percent level for the $\Xi_b$.

In addition to the strangeness changing decays of heavy cascade hyperons
with emission of a pion, we consider here, as a `by-product', two other
types of $\Xi_Q \to \Lambda_Q$ transitions: with emission of a lepton
pair and with emission of a photon, which both of course have much lower
probability than the pion transitions. For the semileptonic transitions
we note that the axial form factor of the weak current is zero in the
heavy quark limit for both $\Xi_b \to \Lambda_b \, \ell \, \nu$ and
$\Xi_c \to \Lambda_c \, \ell \, \nu$, while the vector form factor is
$g_V=1$ in the flavor SU(3) limit (with the usual further consequences
of the Ademollo-Gatto theorem for the SU(3) violation effects). For the
photon transitions it is concluded here that the decay $\Xi_b \to
\Lambda_b \, \gamma$ is forbidden in the heavy quark limit, while this
generally is not the case for $\Xi_c \to \Lambda_c \, \gamma$.

The discussed transitions among the heavy hyperons are induced by two
underlying weak processes: the `spectator' decay of a strange quark,
$s \to u \, \overline u \, d$, $s \to u \, \ell \, \nu$, or $s \to d \,
\gamma$, which does not involve the heavy quark, and the `non-spectator'
weak scattering (WS)
\beq
s \, c \to c \, d
\label{ws}
\eeq
trough the weak interaction of the $c \to d$ and $s \to c$ currents.
One can also readily see that the WS mechanism contributes only to the
decays $\Xi_c \to \Lambda_c \, \pi$ and generally, through a photon
emission in WS, to the radiative transition $\Xi_c^+ \to \Lambda_c \,
\gamma$, while the semileptonic decay $\Xi_c^0 \to \Lambda_c \, \ell \,
\nu$ and all the transitions between the $\Xi_b$ hyperons and the
$\Lambda_b$ are contributed only by the `spectator' processes.

An important starting point in considering the transitions $\Xi_Q \to
\Lambda_Q$ induced by the `spectator' decay of the strange quark, is
that in the heavy quark limit the spin of the heavy quark completely
decouples from the spin variables of the light component of the baryon,
and that the latter light component in both the initial and the final
baryon forms a $J^P = 0^+$ state with the quantum numbers of a diquark.
Thus these transitions are of a $0^+ \to 0^+$ type, which imposes strong
constraints on the decay amplitudes. In particular, for the pion
emission, these constraints imply that the decay amplitude is purely $S$
wave, while the $P$ wave amplitude is zero in the limit of infinite
heavy quark mass\footnote{For a general phenomenological treatment of
the amplitudes of hyperon pion transitions, $B^\prime \to B \, \pi$, in
terms of partial waves, see e.g. the textbook \cite{lbo}.}. The
implication for the `spectator' radiative transition is that it is
forbidden in this limit, thus predicting a strong suppression of the
decay $\Xi_b^0 \to \Lambda_b \, \gamma$. For the semileptonic decays,
contributed only by the `spectator' decay, the constraint from the  $0^+
\to 0^+$ transition is that the axial hadronic form factor is vanishing,
$g_A=0$, while the vector form factor is $g_V=1$. The corresponding
decay rate $\Gamma (\Xi_Q \to \Lambda_Q \, e \, \nu) = G_F^2 \, \sin^2
\theta_c \, (\Delta M)^5/(60 \pi^3) \approx 1.0 \times 10^6 \, s^{-1}$
is however too small to be of a possible phenomenological relevance in
the nearest future.

For further consideration of the pion transitions $\Xi_Q \to \Lambda_Q
\, \pi$ we write the well known expression for the nonleptonic
strangeness-changing weak Hamiltonian (see e.g. \cite{lbo})
\begin{eqnarray}
H_W = &&\sqrt{2} \, G_F \, \sin \theta_c \left \{ \left ( C_+ + C_-
\right ) \, \left [ (\overline u_L \, \gamma_\mu \, s_L)\, (\overline
d_L \, \gamma_\mu \, u_L) - (\overline c_L \, \gamma_\mu \, s_L)\,
(\overline d_L \, \gamma_\mu \, c_L) \right ] + \right . \nonumber \\
&& \left. \left ( C_+ - C_- \right ) \, \left [ (\overline d_L \,
\gamma_\mu \, s_L)\, (\overline u_L \, \gamma_\mu \, u_L) - (\overline
d_L \, \gamma_\mu \, s_L)\, (\overline c_L \, \gamma_\mu \, c_L) \right
] \right \}~.
\label{hw}
\end{eqnarray}
In this formula the weak Hamiltonian is assumed to be normalized (in
LLO) at $\mu = m_c$, so that the renormalization coefficients are
$C_-=C_+^{-2}= \left ( \alpha_s(m_c)/\alpha_s(m_W) \right )^{12/25}$.
The terms in the Hamiltonian (\ref{hw}) without the charmed quark fields
describe the `spectator' nonleptonic decay of the strange quark, while
those with the $c$ quark correspond to the WS process (\ref{ws}). It
should be noted that the part of $H_W$ with (virtual) charmed quarks
indirectly contributes to the `spectator' process as well by providing a
GIM cutoff for the `penguin' mechanism \cite{svz} at $\mu=m_c$. However
at any normalization point $\mu$ below $m_c$ this part does not
explicitly show up, and for this reason we refer to the terms of $H_W$
without the charmed quark fields as `spectator' ones and those with $c$
and $\overline c$ as `non-spectator' ones. One could evolve the weak
Hamiltonian down to a low normalization point $\mu$, such that $\mu \ll
m_c$ to make this separation explicit (the `spectator' part then evolves
according to the treatment in Ref.\cite{svz}, while the evolution of the
`non-spectator' part is described by the `hybrid' anomalous dimension
\cite{sv}, and is essentially equivalent to that considered in
\cite{sv3}). However, the present paper makes no attempt at constructing
models for the hadronic matrix elements at a low $\mu$, thus writing the
corresponding formulas here would be redundant and it is quite
sufficient for our purposes to use the expression (\ref{hw}) at
$\mu=m_c$ with the separation between the `spectator' and the
`non-spectator' parts as noted.

As discussed above, the `spectator' process gives rise only to the $S$
wave amplitudes of the decays $\Xi_Q \to \Lambda_Q \, \pi$, while the
`non-spectator' part involves the spin of the charmed quark, and
generally may induce a $P$ wave as well as an $S$ wave in the decays of
the $\Xi_c$ hyperons.  According to the well known current algebra
technique, the $S$ wave amplitudes of pion emission can be considered in
the chiral limit at zero four-momentum of the pion, where they are
described by the PCAC reduction formula (pole terms are absent in these
processes):
\beq
\langle \Lambda_Q \, \pi_i (p=0) \,| H_W |\, \Xi_Q \rangle = {\sqrt{2}
\over f_\pi} \, \langle \Lambda_Q \, |\left[Q^5_i, \, H_W \right ] |\,
\Xi_Q \rangle~,
\label{pcac}
\eeq
where $\pi_i$ is the pion triplet in the Cartesian notation, and $Q^5_i$
is the corresponding isotopic triplet of axial charges. The constant
$f_\pi \approx 130 \, MeV$, normalized by the charged pion decay, is
used here, hence the coefficient $\sqrt{2}$ in eq.(\ref{pcac}).

It is straightforward to see from eq.(\ref{pcac}) that in the PCAC limit
the discussed decays should obey the $\Delta I =1/2$ rule. Indeed, the
commutator of the weak Hamiltonian with the axial charges transforms
under the isotopic SU(2) in the same way as the Hamiltonian itself. In
other words, the $\Delta I=1/2$ part of $H_W$ after the commutation
gives an $\Delta I=1/2$ operator, while the $\Delta I = 3/2$ part after
the commutation gives an $\Delta I = 3/2$ operator. The latter operator
however cannot have a non vanishing matrix element between an isotopic
singlet, $\Lambda_Q$, and an isotopic doublet, $\Xi_Q$. Thus the $\Delta
I=3/2$ part of $H_W$ gives no contribution to the $S$ wave amplitudes in
the PCAC limit. The $P$ wave part of the amplitude vanishes at a zero
pion momentum and thus is not given by eq.(\ref{pcac}). However, as
discussed, the $P$ wave can arise only in the decays of charmed hyperons
and only from the `non-spectator' part of the $H_W$, which is pure
$\Delta I =1/2$ to start with.

Once the isotopic properties of the decay amplitudes are fixed, one can
concentrate on specific charge decay channels, e.g. $\Xi_b^- \to
\Lambda_b \, \pi^-$ and $\Xi_c^0 \to \Lambda_c \, \pi^-$. An application
of the PCAC relation (\ref{pcac}) with the Hamiltonian from
eq.(\ref{hw}) to these decays, gives the expressions for the amplitudes
at $p=0$ in terms of baryonic matrix elements of four-quark operators:
\begin{eqnarray}
&&\langle \Lambda_b \, \pi^- (p=0) \,| H_W |\, \Xi_b^- \rangle =
\nonumber \\
&&{\sqrt{2} \over f_\pi} \, G_F \, \sin \theta_c \, \langle \Lambda_b
\,| \left ( C_+ + C_- \right ) \, \left [ (\overline u_L \, \gamma_\mu
\, s_L)\, (\overline d_L \, \gamma_\mu \, d_L) - (\overline u_L \,
\gamma_\mu \, s_L)\, (\overline u_L \, \gamma_\mu \, u_L) \right ] +
\nonumber \\
&& \left ( C_+ - C_- \right ) \, \left [ (\overline d_L \, \gamma_\mu \,
s_L)\, (\overline u_L \, \gamma_\mu \, d_L) - (\overline u_L \,
\gamma_\mu \, s_L)\, (\overline u_L \, \gamma_\mu \, u_L) \right] | \,
\Xi_b^- \rangle = \nonumber \\
&&{\sqrt{2} \over f_\pi} \, G_F \, \sin \theta_c \, \langle \Lambda_b
\,| C_- \, \left [ (\overline u_L \, \gamma_\mu \, s_L)\, (\overline d_L
\, \gamma_\mu \, d_L) - (\overline d_L \, \gamma_\mu \, s_L)\,
(\overline u_L \, \gamma_\mu \, d_L) \right ] - \nonumber \\
&& {C_+ \over 3} \left [ (\overline u_L \, \gamma_\mu \, s_L)\,
(\overline d_L \, \gamma_\mu \, d_L) + (\overline d_L \, \gamma_\mu \,
s_L)\, (\overline u_L \, \gamma_\mu \, d_L) +2 \, (\overline u_L \,
\gamma_\mu \, s_L)\, (\overline u_L \, \gamma_\mu \, u_L) \right ] | \,
\Xi_b^- \rangle~,
\label{xib}
\end{eqnarray}
where in the last transition the operator structure with $\Delta I =3/2$
giving a vanishing contribution is removed and only the structures with
explicitly $\Delta I =1/2$ are retained, and
\begin{eqnarray}
&&\langle \Lambda_c \, \pi^- (p=0) \,| H_W |\, \Xi_c^0 \rangle = \langle
\Lambda_b \, \pi^- (p=0) \,| H_W |\, \Xi_b^- \rangle +  \nonumber \\
&&{\sqrt{2} \over f_\pi} \, G_F \, \sin \theta_c \, \langle \Lambda_c \,
| \left ( C_+ + C_- \right ) \, (\overline c_L \, \gamma_\mu \, s_L)\,
(\overline u_L \, \gamma_\mu \, c_L) + \nonumber \\
&&\left ( C_+ - C_- \right ) \, (\overline u_L \, \gamma_\mu \, s_L)\,
(\overline c_L \, \gamma_\mu \, c_L) |\, \Xi_c^0 \rangle~.
\label{xic}
\end{eqnarray}
In the latter formula the first term on the r.h.s. expresses the fact
that in the heavy quark limit the `spectator' amplitudes do not depend
on the flavor or the mass of the heavy quark\footnote{The
non-relativistic normalization for the heavy quark field is used here,
corresponding to $\langle Q | Q^\dagger Q | Q \rangle=1$. Thus the
amplitudes do not contain normalization factors related to the heavy
quark mass.}. The rest of the expression (\ref{xic}) describes the
`non-spectator' contribution to the $S$ wave of the charmed hyperon
decay. Using the flavor SU(3) symmetry this contribution can be related
to the difference of lifetimes within the charmed hyperon triplet as
follows.

Due to the absence of correlation of the spin of the heavy quark in the
hyperons with its light `environment',  the terms, involving the axial
current of the charmed quark in the operators in eq.(\ref{xic}), give no
contribution to the matrix elements. Thus the only relevant matrix
elements are
\beq
\langle \Lambda_c \, | (\overline c \, \gamma_\mu \, c) (\overline u \,
\gamma_\mu \, s) | \, \Xi_c^0 \rangle =-x~~~
%\label{mx}
%\eeq
{\rm and}~~~
%\beq
\langle \Lambda_c \, | (\overline c_i \, \gamma_\mu \, c_k) (\overline
u_k \, \gamma_\mu \, s_i) | \, \Xi_c^0 \rangle =-y~,
\label{mxy}
\eeq
where, by the SU(3) symmetry, the quantities $x$ and $y$ coincide with
those introduced in \cite{mv2} in terms of differences of diagonal
matrix elements over the hyperons:
\begin{eqnarray}
\label{defxy}
x=\left \langle  {1 \over 2} \, (\overline c \, \gamma_\mu \, c) \left
[
(\overline u \, \gamma_\mu u) - (\overline s \, \gamma_\mu s) \right]
\right \rangle_{\Xi_c^0-\Lambda_c} = \left \langle  {1 \over 2} \,
(\overline c \, \gamma_\mu \, c) \left [ (\overline s \, \gamma_\mu s) -
(\overline d \, \gamma_\mu d) \right] \right \rangle_{\Lambda_c -
\Xi_c^+}~,  \\ \nonumber
y=\left \langle  {1 \over 2} \, (\overline c_i \, \gamma_\mu \, c_k)
\left [ (\overline u_k \, \gamma_\mu u_i) - (\overline s_k \, \gamma_\mu
s_i) \right ] \right \rangle_{\Xi_c^0-\Lambda_c} = \left \langle  {1
\over 2} \, (\overline c_i \, \gamma_\mu \, c_k) \left [ (\overline s_k
\, \gamma_\mu s_i) - (\overline d_k \, \gamma_\mu d_i) \right] \right
\rangle_{\Lambda_c - \Xi_c^+}
\end{eqnarray}
with the notation for the differences of the matrix elements:
$\langle {\cal O} \rangle_{A-B}= \langle A | {\cal O} | A \rangle -
\langle B | {\cal O} | B \rangle$.

In terms of the quantities $x$ and $y$ in eq.(\ref{mxy}) the difference
of the $S$ wave decay amplitudes from eq.(\ref{xic}) is written as
\begin{eqnarray}
\Delta A_S \equiv \langle \Lambda_c \, \pi^- (p=0) \,| H_W |\, \Xi_c^0
\rangle - \langle \Lambda_b \, \pi^- (p=0) \,| H_W |\, \Xi_b^- \rangle =
\nonumber \\
{G_F \sin \theta_c \over 2 \, \sqrt{2} \, f_\pi} \, \left[ \left ( C_- -
C_+ \right ) \, x - \left ( C_+ + C_- \right ) \, y \right ]~.
\label{das}
\end{eqnarray}
On the other hand, the same quantities defined by eqs.(\ref{defxy})
describe within the heavy quark expansion \cite{sv3} the differences of
the inclusive weak decay rates within the triplet of the charmed baryons
\cite{mv}. These quantities were extracted \cite{mv} from the current
data on the lifetime differences for the charmed baryons. In particular
it was found that the naive quark model relation $x=-y$ between the
($\mu$ independent) $x$ and the ($\mu$ dependent $y$) does not hold at
any $\mu$ below $m_c$. The numerical value of $x$ is found as $x=-(0.04
\pm 0.01) \, GeV^3$, while the value of $y$ at $\mu=m_c$ is found to be
$y=0.019 \pm 0.009 \, GeV^3$, which values can be directly used in
eq.(\ref{das}). Because of a correlation in the errors in these
estimates and for possible future reference to (hopefully) more precise
future data on the lifetimes, it is rather appropriate to express the
difference of the amplitudes $\Delta A_S$ in eq.(\ref{das}) directly in
terms of the lifetimes of the charmed hyperons, using the formulas from
Ref.\cite{mv}. In terms of the operators normalized at $\mu=m_c$ the
relations for the differences of the total decay rates, including the
dominant Cabibbo-unsuppressed nonleptonic decays as well as the decays
with single Cabibbo suppression and the semileptonic decays, read as
\begin{eqnarray}
&&\Gamma(\Xi_c^0)-\Gamma(\Lambda_c)={G_F^2 \, m_c^2 \over 4 \pi} \,
\cos^2 \theta_c \, \left \{ -x \, \left[ \cos^2 \theta_c \, C_+ \, C_-
+{\sin^2 \theta_c \over 4} \, \left ( 6 \, C_+ \, C_- + 5 \, C_+^2 + 5
\, C_-^2 \right ) \right ] + \right. \nonumber \\
&& \left.
y \, \left [ 3 \, \cos^2 \theta_c \, C_+ \, C_- + {3 \,\sin^2 \theta_c
\over 4} \, \left ( 6 \, C_+ \, C_- -3 \, C_+^2 + C_-^2 \right ) +2
\right ] \right \} \approx
{G_F^2 \, m_c^2 \over 4 \pi} \, \left ( -1.39 \, x + 5.56 \, y \right)~,
\nonumber \\
\label{difgam}
&&\Gamma(\Lambda_c)-\Gamma(\Xi_c^+)={G_F^2 \, m_c^2 \over 4 \pi} \,
\left \{ -x \,  {\cos^4 \theta_c \over 4}\, \left ( 5 \, C_+^2 +5 \,
C_-^2 -2 \, C_+ \, C_-  \right ) + \right. \\ \nonumber
&& \left.
y \, \left [ {3 \, \cos^4 \theta_c \over 4} \, \left (  C_-^2 -3 \,
C_+^2 - 2 \, C_+ \, C_- \right ) - 2 \left ( \cos^2 \theta_c - \sin^2
\theta_c \right) \right ] \right \} \approx -{G_F^2 \, m_c^2 \over 4
\pi} \, \left( 2.88 \, x + 3.16 \, y \right )~,
\end{eqnarray}
where for simplification of the subsequent relation the numerical values
are substituted for the coefficients $C_+$ and $C_-$: $C_+ =0.80$,
$C_-=1.55$, corresponding to $\alpha_s(m_c)/\alpha_s(m_W)=2.5$.

The relations (\ref{difgam}) allow to eliminate the quantities $x$ and
$y$ from the expression (\ref{das}) in favor of the differences of the
total decay rates:
\begin{eqnarray}
\Delta A_S \approx - {\sqrt{2} \, \pi \, \sin \theta_c \over G_F \,
m_c^2 \, f_\pi} \, \left [ 0.45 \, \left (
\Gamma(\Xi_c^0)-\Gamma(\Lambda_c) \right ) + 0.04 \left (
\Gamma(\Lambda_c)-\Gamma(\Xi_c^+) \right ) \right ] = \nonumber \\
-10^{-7} \left [
0.97 \left ( \Gamma(\Xi_c^0)-\Gamma(\Lambda_c) \right ) + 0.09 \left (
\Gamma(\Lambda_c)-\Gamma(\Xi_c^+) \right ) \right ] \, \left ( {1.4 \,
GeV \over m_c} \right )^2 \, ps ~,
\label{dasm}
\end{eqnarray}
where, clearly, in the latter form the widths are assumed to be
expressed in $ps^{-1}$, and $m_c=1.4 \, GeV$ is used as a `reference'
value for the charmed quark mass. It is seen from eq.(\ref{dasm}) that
the evaluation of the difference of the amplitudes within the discussed
approach is mostly sensitive to the difference of the decay rates of
$\Xi_c^0$ and $\Lambda_c$, with only very little sensitivity to the
total decay width of $\Xi_c^+$. Using the current data \cite{pdg} on the
total decay rates: $\Gamma(\Lambda_c)=4.85 \pm 0.28 \, ps^{-1}$,
$\Gamma(\Xi_c^0)= 10.2 \pm 2 \, ps^{-1}$, and the updated value
\cite{pdg1} $\Gamma(\Xi_c^+)=3.0 \pm
0.45 \, ps^{-1}$, the difference $\Delta A_S$ is estimated as
\beq
\Delta A_S= -(5.4 \pm 2) \times 10^{-7}~,
\label{dasn}
\eeq
with the uncertainty being dominated by the experimental error in the
lifetime of $\Xi_c^0$. An $S$ wave amplitude $A_S$ of the magnitude,
given by the central value in eq.(\ref{dasn}) would produce a decay rate
$\Gamma (\Xi_Q \to \Lambda_Q \, \pi) = |A_S|^2 \, p_\pi/(2 \pi) \approx
0.9 \times 10^{10} \, s^{-1}$, which result can also be written in a
form of triangle inequality
\beq
\sqrt{\Gamma(\Xi_b^- \to \Lambda_b \, \pi^-)} + \sqrt{\Gamma(\Xi_c^0 \to
\Lambda_c \, \pi^-)} \ge \sqrt{0.9 \times 10^{10} \, s^{-1}}~.
\label{tri}
\eeq

Although at present it is not possible to evaluate in a reasonably model
independent way the matrix element in eq.(\ref{xib}) for the `spectator'
decay amplitude, the estimate (\ref{tri}) shows that at least some of
the discussed pion transitions should go at the level of $0.01 \,
ps^{-1}$. Also the numerical result for $\Delta A_S$ invites one more
remark, related to the problem of the differences of lifetimes of heavy
hadrons. Namely, we have seen here that the values of the matrix
elements $x$ and $y$ extracted from the data on the lifetimes of charmed
hyperons, translate, in terms of the pion transitions, into a `natural'
magnitude of the decay amplitude, which one would expect, based on the
experience with the decays of the ordinary strange hyperons. On the
other hand, these phenomenological numerical values of $x$ and $y$ are
deemed to be substantially enhanced with respect to the simple estimates
\cite{sv2,sv3}, $y=-x=f_D^2 \, m_D /12 \approx 0.006 \, GeV^3$, based on
non-relativistic-like ideas about the structure of the heavy baryons and
mesons. The discussed here relation of these matrix elements to the
decays $\Xi_Q \to \Lambda_Q \, \pi$ and the evaluation of their
significance, perhaps, tell us that the phenomenological values of $x$
and $y$ \cite{mv} are of a `normal' magnitude, while the early
simplistic theoretical estimates were simply too low.

To summarize. It is shown that the strangeness-changing decays $\Xi_Q
\to \Lambda_Q \, \pi$ of the $b$ and $c$ cascade hyperons should go at
the rate of order $0.01 \, ps^{-1}$ and that they should obey the
$\Delta I=1/2$ rule. The decays $\Xi_b \to \Lambda_b \, \pi$ are purely
$S$ wave in the heavy quark limit and their amplitude, as well as the
$S$ wave amplitude of the decays $\Xi_s \to \Lambda_s \, \pi$, is
expressed by the current algebra in terms of matrix elements of
four-quark operators over the heavy baryons. The difference between the
$S$ wave amplitude of the charmed hyperon decay and that of the $\Xi_b$
is related, within the heavy quark expansion, to the differences of
lifetimes among the heavy hyperons. The semileptonic transitions $\Xi_Q
\to \Lambda_Q \, \ell \, \nu$ are of a purely $0^+ \to 0^+$ type and
thus have $g_A=0$, $g_V=1$, while the radiative weak decay $\Xi_b \to
\Lambda_b \, \gamma$ is shown to be greatly suppressed in the heavy
quark limit. \\[0.2in]

\noindent
{\large \bf Acknowledgement}

\noindent
This work is supported in part by DOE under the grant number
DE-FG02-94ER40823.


\begin{thebibliography}{99}

\bibitem{pdg}
Particle Data Group, Eur. Phys. J. {\bf C 3} (1998) 1.
\bibitem{bigi}
I.I. Bigi, Univ. Notre Dame report UND-HEP-99-BIG 07, Jan. 2000; \
[hep-ph/0001003].
\bibitem{gms}
B. Guberina, B. Meli\'{c} and H. \v Stefan\v ci\'c, \pl{B469}{99}{253}.
\bibitem{mv}
M.B. Voloshin, Univ. Minnesota TPI report TPI-MINN-99/40-T, Aug. 1999. \
[hep-ph/9908455]. Phys. Rev. D to be published.
\bibitem{lbo}
L.B. Okun, {\it Leptons and Quarks}, North-Holland, Amsterdam, 1982,
1984.
\bibitem{svz}
A.I. Vainshtein, V.I. Zakharov and M.A. Shifman, ZhETF {\bf 72} (1977)
1275; \spj{45}{77}{670}.
\bibitem{sv}
M.A. Shifman and M.B. Voloshin, Yad.Fiz. {\bf 45} (1987) 463;
\sjnp{45}{87}{292}.
\bibitem{sv3}
M.A. Shifman and M.B. Voloshin, \spj{64}{86}{698}.
\bibitem{mv2}
M.B. Voloshin, \prep{320}{99}{275}.
\bibitem{pdg1}
Particle Data Group. {\it 1999 WWW Update},
http://pdg.lbl.gov/1999/bxxx.html
\bibitem{sv2}
M.A. Shifman and M.B. Voloshin, \sjnp{41}{85}{120}.





\end{thebibliography}
\end{document}